\documentclass[11pt]{article}
\usepackage{graphicx}
\usepackage{latexsym,graphicx}
\usepackage{float}
\usepackage{amsmath}
\usepackage{amsfonts}
\usepackage{amssymb}
\usepackage{epstopdf}
\usepackage{xcolor}
\DeclareGraphicsExtensions{.eps}
\usepackage[left=2.5cm,top=2.5cm,right=2.5cm,bottom=1.5cm]{geometry}

\catcode `\@=11
\catcode `\@=12
\begin{document}
	\begin{center}
		\large{\bf{Brans-Dicke Scalar Field Cosmological Model in Lyra's Geometry}} \\
		\vspace{5mm}
		\normalsize{Dinesh Chandra Maurya$^1$, Rashid Zia$^2$}\\
		\vspace{5mm}
				\normalsize{$^{1}$Department of Mathematics, Faculty of Engineering \& Technology, IASE (Deemed to be University), Sardarshahar-331 403, Rajsthan, India \\
			\vspace{2mm}
		\normalsize{.$^{2}$Department of Mathematics, BBS College of Engineering and Technology, Allahabad, U.P., 211003, India}\\
		\vspace{2mm}
			$^1$E-mail:dcmaurya563@gmail.com \\
			\vspace{2mm}
			$^2$E-mail:rashidzya@gmail.com}\\
	\end{center}
	\vspace{5mm}
	\begin{abstract}
		In this paper, we have developed a new cosmological model in Einstein's modified gravity theory using two types of modification.(i) Geometrical modification, in which we have used Lyra's geometry in the left hand side of the Einstein field equations (EFE) and (ii) Modification in gravity (energy momentum tensor) on right hand side of EFE, as per Brans-Dicke (BD) model. With these two modifications, we have investigated a spatially homogeneous and anisotropic Bianchi type-I cosmological models of Einstein's Brans-Dicke theory of gravitation in Lyra geometry. The model represents accelerating universe at present and decelerating in past and is considered to be dominated by dark energy. Gauge function $\beta$ and BD-scalar field $\phi$ are considered as a candidate for the dark energy and is responsible for the present acceleration. The derived model agrees at par with the recent supernovae (SN Ia) observations. We have set BD-coupling constant $\omega$ to be greater than 40000, seeing the solar system tests and evidences. We have discussed the various physical and geometrical properties of the models and have compared them with the corresponding relativistic models.
	\end{abstract}
	\smallskip
	\vspace{5mm}
	\smallskip
	\section{Introduction}
	Einstein formulated the theory of General Relativity (GR) in 1915, in which, he described gravity as a geometrical property of space and time. In particular, the curvature of space-time was proposed to be, directly related to the energy and momentum of whatever matter and radiation are present in the universe. The relation was specified by the Einstein field equations (EFE), which are a system of partial differential equations. The original EFE were written in the form $ R_{\mu\nu} - \frac{1}{2}Rg_{\mu\nu} = \frac{8{\pi} G}{c^4} T_{\mu\nu} $, where $R_{\mu\nu} $ is Ricci curvature tensor, R is the scalar curvature, $g_{\mu\nu} $ is the metric tensor, and $T_{\mu\nu} $ is the energy momentum tensor.
	\par EFE are a foundation on which various cosmological models have been constructed. Soon after formulation of the field equation, Einstein himself, applied these equations for constructing the model of the universe. Einstein's original field equations support an expanding universe. But at that time, it was believed that the universe is static. So to make his model static, Einstein himself modified his original EFE by introducing a positive constant $\Lambda$, in his field equations, and termed this as cosmological constant. But, after Hubble's discovery in 1929, it was established that universe is not static but expanding. So, under the new scenario, Einstein abandoned the cosmological constant, calling it the "biggest blunder" of his life, and back to his original field equations. Inspired by geometrizing gravitation, in 1918, Weyl \cite{ref1} proposed a more general theory in which both gravitation and electromagnetism are described geometrically.
	\par So, to accommodate new findings or for the sake of generalization, various modifications have been proposed by researchers in original EFE from time to time, since it's inception. Some researchers modified the geometrical part, whereas some proposed the modification in energy momentum part of the EFE.
	\par In the present paper, we have applied both types of modification simultaneously. On the left hand side, we have used Lyra's geometry (Geometrical modification) in EFE and on the right hand side we have modified energy momentum tensor as per Brans-Dicke model. The motivation behind such modifications are  as follows:\\ 
	In general relativity Einstein succeeded in geometrizing gravitation by identifying the metric tensor with the gravitational potentials. 
	In 1951, Lyra proposed a modification of Riemannian geometry by introducing a gauge function into the structureless manifold \cite{ref2}, (see also \cite{ref3}). Based on Lyra geometry, a new scalar-tensor theory of gravitation which was an analogue of the Einstein field equations was proposed by Sen and Dunn (\cite{ref4,ref5}). An interesting feature of this model is that, it keeps the spirit of Einstein’s principle of geometrization, since both the scalar and tensor fields have more or less intrinsic geometrical significance. In contrast, in the Brans-Dicke theory, the tensor field alone is geometrized and the scalar field remains alien to the geometry (\cite{ref6,ref7}). By incorporating both of these separate modifications in a single theory, we have arrived at a more general method of geometrizing gravitation.
	\par Also, in the construction of recent cosmological models, two problems prevail. The late time acceleration problem and the existence of big bang singularity. To deal with the first problem, one of the way is, by introducing the gauge function into the structureless manifold in the framework of Lyra geometry. In this approach, the cosmological constant of EFE naturally arises from the geometry, instead of introducing it in an ad-hoc way. In fact, the constant displacement vector field plays the role of cosmological constant which can be responsible for the late-time cosmological acceleration of universe (\cite{ref8,ref9}). For the second problem, recently, a new mechanism for avoiding the big-bang singularity was proposed in the framework of the emergent universe scenario (see \cite{ref10,ref11,ref12}). The emergent universe scenario is a past-eternal inflationary model in which the horizon problem is solved before the beginning of inflation and the big-bang singularity is removed.
	\par Till date, various researchers have studied cosmology in Lyra's geometry (see, \cite{ref8,ref9}, \cite{ref13}-\cite{ref29}) with both, a constant displacement field and a time-dependent one. For instance, in \cite{ref20} the displacement field is allowed to be time dependent, and the Friedmann-Robertson-Walker (FRW) models are derived in Lyra's manifold. Those models are free of the big-bang singularity and solve the entropy and horizon problems which beset the standard models based on Riemannian geometry. Recently, cosmological models in the frame work of Lyra's geometry in different contexts are investigated in several papers (see, \cite{ref21}-\cite{ref29}).
	\par Also, in last few decades, there has been considerable interest in alternative theories of gravitation coursed by the investigations of inflation and, especially, late cosmological acceleration which is well proved in many papers (see \cite{ref30}-\cite{ref35}). In order to explain such unexpected behavior of our universe, one can modify the gravitational theory (see \cite{ref36}-\cite{ref41}), or construct various field models of the so-called dark energy (DE), for which equation of state (EoS) satisfies $\gamma = \frac{p}{\rho} < -\frac{1}{3}$.(see \cite{ref42}-\cite{ref48}). Presently, there is an uprise of interest in scalar fields in GR and alternative theories of gravitation in this context. Therefore, the study of cosmological scalar-field models in Lyra's geometry is relevant for the cosmic acceleration models. A Bianchi type-I dust filled accelerating Brans-Dicke cosmological model with cosmological constant $\Lambda$ as a dark energy candidate was investigated by Goswami et al. \cite{ref54} and a Brans-Dicke scalr-field cosmological models in Lyra geometry with time dependent deceleration parameter was studied by \cite{ref61}. Recently, a detailed review on dark energy/modified gravity problem was presented by \cite{ref66}.\\
	\par Most studies in Lyra's cosmology involve a perfect fluid. Strangely, at least up to our knowledge, the case of scalar field in Lyra's cosmology was not studied properly. Here we would like to fill this gap. In this paper, we will consider a scalar field Brans-Dicke cosmology in the context of Lyra's geometry. With motivation provided above, we have investigated Einstein's modified field equations for the spatially homogeneous anisotropic Bianchi Type-I space-time metric within the frame work of Lyra's geometry.
	\par The out line of the paper is as follows: section-$1$ is introductory in nature. In section-$2$, the field equations in Lyra geometry with Brans-Dicke modifications are described. Section-$3$ deals with	the cosmological solutions and have established relations among energy parameters $\Omega_{m},~\Omega_{\sigma}$ and $\Omega_{\beta}$. In section-$4$, we obtained expressions for Hubble's constant, luminosity distance and apparent magnitude in terms of red shift and scale factor. We have also estimated the present values of energy parameters and Hubble's constant. The deceleration parameter (DP), age of the universe and certain physical properties of the universe are presented in section-$5$. The discussion of results are given in section-$6$. Finally, conclusions are summarized in section-$7$. \\
	
	\smallskip
	\vspace{5mm}
	\section{Lyra Geometry and Einstein's Brans-Dicke field equations}
	Lyra geometry is a modification of Riemannian geometry by introducing a gauge function into the structureless
	manifold \cite{ref2}, see also \cite{ref3}. Lyra defined a displacement vector between two neighboring points $P(x^{\mu})$
	and $Q(x^{\mu} + dx^{\mu})$ as $Adx^{\mu}$ where $A = A(x^{\mu})$ is a non-zero gauge function of the coordinates. The gauge
	function $A(x^{\mu})$ together with the coordinate system $x^{\mu}$ form a reference system $(A, x^{\mu})$. The transformation
	to a new reference system $(\bar{A}, \bar{x}^{\mu})$ is given by the following functions	
	\begin{align}\label{1}
	\bar{A}=\bar{A}(A, x^{\mu}), \hspace{1cm}  \bar{x}^{\mu}=\bar{x}(x^{\mu}),
	\end{align}
	where $\frac{\partial\bar{A}}{\partial A}\neq 0$ and $det\left(\frac{\partial\bar{x}}{\partial x}\right)\neq 0$.\\
	The symmetric affine connections $\tilde{\Gamma}_{\nu\sigma}^{\mu}$ on this manifold is given by
	\begin{equation}\label{2}
	\tilde{\Gamma}_{\nu\sigma}^{\mu}=\frac{1}{A}\Gamma_{\nu\sigma}^{\mu}+\frac{1}{2}(\delta_{\nu}^{\mu}\psi_{\sigma}+\delta_{\sigma}^{\mu}\psi_{\nu}-g_{\nu\sigma}\psi^{\mu})
	\end{equation}
	where the connection $\Gamma_{\nu\sigma}^{\mu}$ is defined in terms of the metric tensor $g_{\mu\nu}$ as in Riemannian geometry and $\psi^{\mu}=g^{\mu\nu}\psi_{\nu}$ is the so-called displacement vector field of Lyra geometry. It is shown by Lyra \cite{ref2}, and also by
	Sen \cite{ref4}, that in any general reference system, the displacement vector field $\psi^{\mu}$ arises as a natural consequence
	of the formal introduction of the gauge function $A(x^{\mu})$ into the structureless manifold. Equation (2) shows that the component of the affine connection, not only depends on metric $g_{\mu\nu}$ but also on the displacement vector field $\psi^{\mu}$. The line element (metric) in Lyra geometry is given by 
	\begin{equation}\label{3}
	ds^{2}=A^{2}g_{\mu\nu}dx^{\mu}dx^{\nu}
	\end{equation}
	which is invariant under both of the coordinate and gauge transformations. The infinitesimal parallel transport
	of a vector field $V^{\mu}$ is given by
	\begin{equation}\label{4}
	\delta V^{\mu}=\hat{\Gamma}_{\nu\sigma}^{\mu}V^{\nu}Adx^{\sigma}
	\end{equation}
	where $\hat{\Gamma}_{\nu\sigma}^{\mu}=\tilde{\Gamma}_{\nu\sigma}^{\mu}-\frac{1}{2}\delta_{\nu}^{\mu}\psi_{\sigma}$ which is not symmetric with respect to $\nu$ and $\sigma$. In Lyra geometry, unlike Weyl geometry, the connection is metric preserving as in Riemannian geometry which indicates that length transfers are integrable. This means that the length of a vector is conserved upon parallel transports, as in
	Riemannian geometry.\\
	The curvature tensor of Lyra geometry is defined in the same manner as in Riemannian geometry and is given by
	\begin{equation}\label{5}
	\tilde{R}^{\mu}_{\nu\rho\sigma}=A^{-2}\left[\frac{\partial}{\partial x^{\rho}}(A\hat{\Gamma}_{\nu\sigma}^{\mu})-\frac{\partial}{\partial x^{\sigma}}(A\hat{\Gamma}_{\nu\rho}^{\mu})+A^{2}(\hat{\Gamma}_{\lambda\rho}^{\mu}\hat{\Gamma}_{\nu\sigma}^{\lambda}-\hat{\Gamma}_{\lambda\sigma}^{\mu}\hat{\Gamma}_{\nu\rho}^{\lambda}) \right] 
	\end{equation}
	Then, the curvature scalar of Lyra geometry will be
	\begin{equation}\label{6}
	\tilde{R}=A^{-2}R+3A^{-1}\nabla_{\mu}\psi^{\mu}+\frac{3}{2}\psi^{\mu}\psi_{\mu}+2A^{-1}(log A^{2})_{,\mu}\psi^{\mu}
	\end{equation}
	where $R$ is the Riemannian curvature scalar and the covariant derivative is taken with respect to the Christoffel
	symbols of the Riemannian geometry.\\
	The invariant volume integral in four dimensional Lyra manifold is given by
	\begin{equation}\label{7}
	I=\int\sqrt{-g}L(Adx)^{4}
	\end{equation}
	where $L$ is an invariant scalar in this geometry. Using the normal gauge $A = 1$ and $L = \tilde{R}$ through the
	equations (6) and (7) results in
	\begin{equation}\label{8}
	\tilde{R}=R+3\nabla_{\mu}\psi^{\mu}+\frac{3}{2}\psi^{\mu}\psi_{\mu}
	\end{equation}	
	\begin{equation}\label{9}
	I=\int\sqrt{-g}\tilde{R}dx^{4}
	\end{equation}		
	Therefore the Lagrangian for Brans-Dicke theory in Lyra geometry can be defined as
	\begin{equation}\label{10}
	\tilde{L}_{BDT}=\phi\left(\tilde{R}-w\frac{\phi_{,\mu}\phi^{,\mu}}{\phi^{2}}\right)+16\pi L_{mat}. 
	\end{equation}
	where $\tilde{R}=R+3\nabla_{\mu}\psi^{\mu}+\frac{3}{2}\psi^{\mu}\psi_{\mu}$ is curvature scalar of Lyra geometry \cite{ref2} using normal gauge transformations and $\psi^{\mu}$ is a displacement vector field of Lyra geometry, $R$ is curvature scalar in Riemannian geometry, $w$ is the Brans-Dicke coupling constant, $\phi$ is Brans-Dicke scalar field as mentioned in the first section and $L_{mat}$ is Lagrangian for matter. Therefore, the action for this Lagrangian is defined as
	\begin{equation}\label{11}
	I=\int\left[\phi\left(R+3\nabla_{\mu}\psi^{\mu}+\frac{3}{2}\psi^{\mu}\psi_{\mu}-w\frac{\phi_{,\mu}\phi^{,\mu}}{\phi^{2}}\right)+16\pi L_{mat}\right]\sqrt{-g}d^{4}x 
	\end{equation}
	By varying the action $I$ of gravitational field with respect to the metric tensor components $g^{\mu\nu}$ and $\phi$ respectively, we obtained the following Einstein's Brans-Dicke field equations in Lyra geometry	
	\begin{equation}\label{12}
	G_{\mu\nu}+\frac{3}{2}\psi_{\mu}\psi_{\nu}-\frac{3}{4}g_{\mu\nu}\psi_{\sigma}\psi^{\sigma}=-\frac{8\pi T_{\mu\nu}}{\phi ~c^{4}}-\frac{w}{\phi^{2}}\left(\phi_{,\mu}\phi_{,\nu}-\frac{1}{2}g_{\mu\nu}\phi_{,\sigma}\phi^{,\sigma} \right)-\frac{1}{\phi}\left( \phi_{,\mu,\nu}-g_{\mu\nu}\Box\phi\right) 
	\end{equation}
	\begin{equation}\label{13}
	\Box\phi=\phi_{,\mu}^{,\mu}=\frac{8\pi T}{(3+2w)~c^{2}}
	\end{equation}
	where $G_{\mu\nu}$ is Einstein curvature tensor, $(,)$ and $(^{,})$ denotes the covariant and contra-variant derivatives and other symbols have their usual meanings in the Riemannian geometry. The metric in Lyra geometry is defined by $ds^{2}=g_{\mu\nu}(Adx^{\mu})(Adx^{\nu})$, where $A$ is the gauge transformations. Using normal gauge transformation $(A=1)$ the above metric becomes $ds^{2}=g_{\mu\nu}dx^{\mu}dx^{\nu}$ as in Riemannian geometry.\\
	We assume a perfect fluid form for the energy-momentum tensor:
	\begin{equation}\label{14}
	T_{\mu\nu}=(\rho+p)u_{\mu}u_{\nu}+pg_{\mu\nu}
	\end{equation}
	and co-moving coordinates $u_{\mu}u^{\mu}=-1$. We also let $\psi_{\mu}$ be the time-like constant vector
	\begin{equation}\label{15}
	\psi_{\mu}=(\beta,0,0,0)
	\end{equation}
	where $\beta$ is a constant. The metric for Bianchi type-I space-time is given by
	\begin{equation}\label{16}
	ds^{2}=-dt^{2}+A^{2}dx^{2}+B^{2}dy^{2}+C^{2}dz^{2}
	\end{equation}
	where $A, B$ and $C$ are functions of cosmic time $t$ alone.\\
	\par For the metric $(16)$, solving the field equations $(14)$, we get the following field equations:
	\begin{equation}\label{17}
	\frac{\dot{A}\dot{B}}{AB}+\frac{\dot{B}\dot{C}}{BC}+\frac{\dot{A}\dot{C}}{AC}-\frac{3}{4}\beta^{2}=\frac{8\pi \rho}{\phi~c^{2}}-\frac{w}{2}\left(\frac{\dot{\phi}}{\phi}\right)^2+\frac{\dot{\phi}}{\phi}\left(\frac{\dot{A}}{A}+\frac{\dot{B}}{B}+\frac{\dot{C}}{C}\right) 
	\end{equation}
	\begin{equation}\label{18}
	\frac{\ddot{B}}{B}+\frac{\ddot{C}}{C}+\frac{\dot{B}\dot{C}}{BC}+\frac{3}{4}\beta^{2}=-\frac{8\pi p}{\phi~c^{2}}+\frac{w}{2}\left(\frac{\dot{\phi}}{\phi}\right)^2+\frac{\dot{\phi}}{\phi}\left(\frac{\dot{B}}{B}+\frac{\dot{C}}{C}\right)+\frac{\ddot{\phi}}{\phi} 
	\end{equation}
	\begin{equation}\label{19}
	\frac{\ddot{A}}{A}+\frac{\ddot{C}}{C}+\frac{\dot{A}\dot{C}}{AC}+\frac{3}{4}\beta^{2}=-\frac{8\pi p}{\phi~c^{2}}+\frac{w}{2}\left(\frac{\dot{\phi}}{\phi}\right)^2+\frac{\dot{\phi}}{\phi}\left(\frac{\dot{A}}{A}+\frac{\dot{C}}{C}\right)+\frac{\ddot{\phi}}{\phi}
	\end{equation}
	\begin{equation}\label{20}
	\frac{\ddot{A}}{A}+\frac{\ddot{B}}{B}+\frac{\dot{A}\dot{B}}{AB}+\frac{3}{4}\beta^{2}=-\frac{8\pi p}{\phi~c^{2}}+\frac{w}{2}\left(\frac{\dot{\phi}}{\phi}\right)^2+\frac{\dot{\phi}}{\phi}\left(\frac{\dot{A}}{A}+\frac{\dot{B}}{B}\right)+\frac{\ddot{\phi}}{\phi}
	\end{equation}
	\begin{equation}\label{21}
	\frac{\ddot{\phi}}{\phi}+\frac{\dot{\phi}}{\phi}\left(\frac{\dot{A}}{A}+\frac{\dot{B}}{B}+\frac{\dot{C}}{C} \right)=\frac{8\pi(\rho-3p)}{(2w+3)\phi~c^{2}} 
	\end{equation}
	Here over dot denotes derivative with respect to time $t$.\\
	\smallskip
	\vspace{5mm}
	\section{Cosmological solutions of the field equations}
	The covariant derivative of the field equation $(12)$ for the Eq.$(14)$, gives the energy conservation law as
	\begin{equation}\label{22}
	\dot{\rho}+3H(\rho+p)=0
	\end{equation}
    The equation of state for the model is defined as
    \begin{equation}\label{23}
    p=\gamma\rho
    \end{equation}
   where $\gamma$ is EoS parameter of the fluid filled in the universe and the Hubble parameter $H$ is given by $H=\frac{\dot{a}}{a}$, where $a$ is average scale factor.\\
    
    \noindent There are two cases for the value of EoS parameter $\gamma$ in equation $(23)$:\\
    
    \noindent \textbf{Case-I:} taking $\gamma=$constant, integrating equation $(22)$, we get
    \begin{equation}\label{24}
    \rho=\rho_{0}(a)^{-3(1+\gamma)}=\rho_{0}(ABC)^{-(1+\gamma)}
    \end{equation}
    \noindent \textbf{Case-II:} taking $\gamma=$variable and assuming $\gamma=\gamma(a)$ in the form 
    \begin{equation}\label{25}
    \gamma=\gamma_{0}+\gamma_{a}(1-a)
    \end{equation}
    where $\gamma_{0}$ is an arbitrary constant and $\gamma_{a}$ is the value of $\left|\frac{d\gamma}{da}\right|_{a=0}$.\\
    
    Using this in Eq. $(22)$, we get
    \begin{equation}\label{26}
    \rho=a^{-3(1+\gamma_{0}+\gamma_{a})}\exp{(\rho_{0}+3a\gamma_{a})}
    \end{equation}
    In the present model, we will consider the case-I assuming $\gamma=$constant.\\
	\par \textbf{Now,} from Eqs. $(18)-(20)$, we obtain
	\begin{equation}\label{27}
	\frac{B}{A}=c_{2}\exp\left( c_{1}\int\frac{\phi}{a^{3}}dt\right) 
	\end{equation}
	\begin{equation}\label{28}
	\frac{C}{A}=c_{4}\exp\left( c_{3}\int\frac{\phi}{a^{3}}dt\right)
	\end{equation}
	\begin{equation}\label{29}
	\frac{C}{B}=c_{6}\exp\left( c_{5}\int\frac{\phi}{a^{3}}dt\right)
	\end{equation}
Now, taking the value of arbitrary integrating constants $c_{2}=c_{4}=c_{6}=1$ and $c_{1}=k, ~c_{3}=-k, ~c_{5}=-2k$ and assuming
\begin{equation}\label{30}
D=\exp\left(\int\frac{k \phi}{(ABC)}dt\right) 
\end{equation}
We get the following relations,
\begin{equation}\label{31}
B=AD ~~~~~~\& ~~~~~~~C=\frac{A}{D}
\end{equation}
The average scale factor $a$ is defined as \textbf{$a=(ABC)^{\frac{1}{3}}$ and using Eq.$(31)$, we obtain}
\begin{equation}\label{32}
a=(ABC)^{\frac{1}{3}}=A
\end{equation}
Therefore, from Eqs. $(17)$ to $(32)$, we obtain
\begin{equation}\label{33}
3\left(\frac{\dot{A}}{A}\right)^{2}- \left(\frac{\dot{D}}{D}\right)^{2}-\frac{3}{4}\beta^{2}=\frac{8\pi \rho}{\phi~c^{2}}-\frac{w}{2}\left(\frac{\dot{\phi}}{\phi}\right)^2+3\frac{\dot{\phi}}{\phi}\left(\frac{\dot{A}}{A}\right) 
\end{equation}
\begin{equation}\label{34}
2\left(\frac{\ddot{A}}{A}\right)+\left(\frac{\dot{A}}{A}\right)^{2}+\left(\frac{\dot{D}}{D}\right)^{2}+\frac{3}{4}\beta^{2}=-\frac{8\pi p}{\phi~c^{2}}+\frac{w}{2}\left(\frac{\dot{\phi}}{\phi}\right)^2+2\frac{\dot{\phi}}{\phi}\left(\frac{\dot{A}}{A}\right)+\frac{\ddot{\phi}}{\phi}
\end{equation}
\begin{equation}\label{35}
\frac{d}{dt}\left(\frac{\dot{D}}{D}\right)+\frac{\dot{D}}{D}\left(3\frac{\dot{A}}{A}-\frac{\dot{\phi}}{\phi}\right)=0
\end{equation}
\begin{equation}\label{36}
\frac{\ddot{\phi}}{\phi}+3\frac{\dot{\phi}}{\phi}\left(\frac{\dot{A}}{A}\right)=\frac{8\pi(\rho-3p)}{(2w+3)\phi~c^{2}}
\end{equation}
\begin{equation}\label{37}
\frac{\dot{\rho}}{\rho}+3(1+\gamma)\frac{\dot{A}}{A}=0
\end{equation}
From Eq. $(31)$, we get
\begin{equation}\label{38}
\frac{\dot{D}}{D}\frac{A^{3}}{\phi}=k ~~~~~\Rightarrow~~\frac{\dot{D}}{D}=\frac{k~\phi}{A^{3}}
\end{equation}

Now, we define the matter energy density parameter $(\Omega_{m})$, curvature anisotropy parameter $(\Omega_{\sigma})$ and dark energy parameter $\Omega_{\beta}$ as
\begin{equation}\label{39}
\Omega_{m}=\frac{8\pi~\rho}{3~c^{2}~H^{2}~\phi}, ~~~~\Omega_{\sigma}=\frac{k^{2}~\phi^{2}}{3~H^{2}A^{6}}, ~~~~\Omega_{\beta}=\frac{\beta^{2}}{4~H^{2}}
\end{equation} 
The deceleration parameter $(q)$ for scale factor and $(q_{\phi})$ for scalar-field are defined as \cite{ref54}
\begin{equation}\label{40}
q=-\frac{\ddot{a}}{a~H^{2}}, ~~~~~~q_{\phi}=-\frac{\ddot{\phi}}{\phi~H^{2}}
\end{equation}
where $a(t)=A$ is average scale factor.\\
Now, using equation $(39)$ in $(33)$, we get
\begin{equation}\label{41}
\Omega_{m}+\Omega_{\sigma}+\Omega_{\beta}=1+\frac{\omega}{6}\xi^{2}-\xi
\end{equation}
Again using Eqs. $(39),~(40)$ in Eq. $(34)$, we get
\begin{equation}\label{42}
\gamma~\Omega_{m}+\Omega_{\sigma}+\Omega_{\beta}=\frac{2}{3}q-\frac{1}{3}+\frac{\omega}{6}\xi^{2}+\frac{2}{3}\xi-\frac{1}{3}q_{\phi}
\end{equation}
Eq. $(36)$ becomes
\begin{equation}\label{43}
-q_{\phi}+3\xi=\frac{3(1-3\gamma)\Omega_{m}}{2\omega+3}
\end{equation}
where $\xi=\frac{\dot{\phi}}{\phi~H}$.
From Eq. $(41)-(43)$, we get
\begin{equation}\label{44}
q-\frac{(\omega-\omega\gamma-3\gamma+2)}{(1-3\gamma)}q_{\phi}+\frac{(3\omega-3\omega\gamma-12\gamma+5)}{(1-3\gamma)}\xi=2
\end{equation} 
This gives the following power law relation between scalar field $\phi$ and scale factor $A$.
\begin{equation}\label{45}
\phi=\phi_{0}\left(\frac{A}{A_{0}}\right)^{\frac{1-3\gamma}{\omega-\omega\gamma-3\gamma+2}} 
\end{equation}
and
\begin{equation}\label{46}
\xi=\frac{1-3\gamma}{\omega-\omega\gamma-3\gamma+2},
\end{equation}
where $A_{0}$ and $\phi_{0}$ are values of scale factor $A$ and scalar field $\phi$ at present. 
Putting this value of $\xi$ in Eq. $(41)$, we get following relationship for energy parameters:
\begin{equation}\label{47}
\Omega_{m}+\Omega_{\sigma}+\Omega_{\beta}=1-\frac{(1-3\gamma)(5\omega-3\omega\gamma-18\gamma+12)}{6(\omega-\omega\gamma-3\gamma+2)^{2}}
\end{equation}
In Brans-Dicke theory as $\omega \to \infty$, we get the following relativistic result:
\begin{equation}\label{48}
\Omega_{m}+\Omega_{\sigma}+\Omega_{\beta}=1
\end{equation}
\subsection{Gravitational Constant Versus Redshift Relation}
As in Brans-Dicke theory, Gravitational constant $G$ is reciprocal of $\phi$ i.e.
\begin{equation}\label{49}
G=\frac{1}{\phi}
\end{equation}
and
\begin{equation}\label{50}
\frac{A_{0}}{A}=1+z
\end{equation}
where $z$ is the redshift.\\
So, from Eqs. $(45),~(49)$ and $(50)$, we obtain
\begin{equation}\label{51}
\frac{G}{G_{0}}=(1+z)^{\frac{1-3\gamma}{\omega-\omega\gamma-3\gamma+2}}
\end{equation}
It is concluded that variation of gravitational constant $G$ over red shift $z$ and coupling constant $\omega$ follows same patterns for both isotropic and anisotropic BD-Universe.\\
This relationship shows that $G$ grows toward the past and in fact it diverges at cosmological singularity. Radar observations, Lunar mean motion and the Viking landers on Mars \cite{ref51} suggest that rate of variation of gravitational constant must be very much slow of order $10^{-12}$ year$^{-1}$. The recent experimental evidence \cite{ref52,ref53} shows that $\omega > 3300$. Accordingly, we consider large coupling constant $\omega$ in this study.\\
From Eqs. $(46)$ and $(49)$, the present rate of gravitational constant is calculated as
\begin{equation}\label{52}
\left( \frac{\dot{G}}{G}\right)_{0}=-\frac{1-3\gamma}{\omega-\omega\gamma-3\gamma+2}H_{0} 
\end{equation}
where $H_{0}\approxeq10^{-12}$ year$^{-1}$.\\
\begin{figure}[H]
	\centering
	\includegraphics[width=10cm,height=8cm,angle=0]{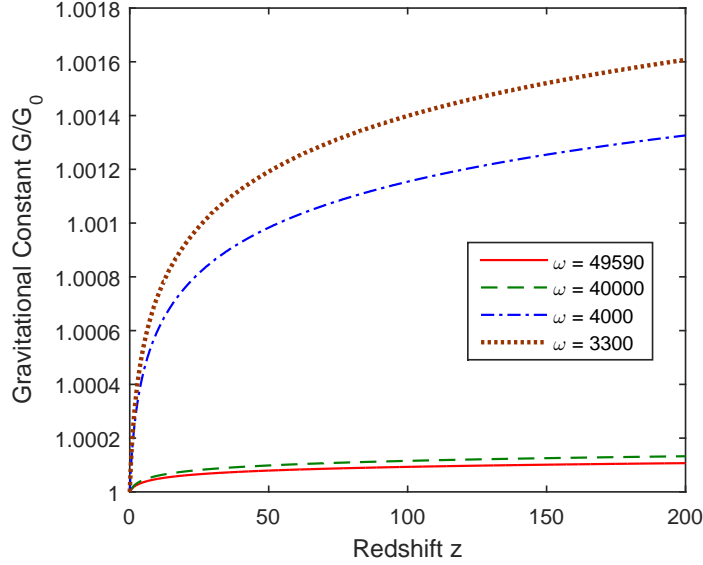}
	\caption{Variation of Gravitational constant over red shift}.
\end{figure}
Eq. $(51)$ exhibits the fact that how $G/G_{0}$ varies over $\omega$. For higher values of $\omega$, $G/G_{0}$ grows very slow over redshift, where as for lower values of $\omega$ it grows fast. The variation of Gravitational constant over redshift for different $\omega$'s are shown in Figure $1$. From Figure $1$ and Eq.$(51)$, it is clear that $G/G_{0}$ and in turn $\frac{1}{\phi}$ is an increasing function of redshift $z$ for $0\leq \gamma \leq\frac{1}{3}$. This implies that $\phi$ is an decreasing function of redshift $z$ and so an increasing function of cosmic time  $t$. It means the value of $G$ decreases with time due to backgrounds effects (i.e. scalar-field). Also, as $\omega\to\infty$, $G/G_{0}\to 1$ i.e. $G=G_{0}$ at $\omega=\infty$ and in this case model behaves like an Einstein's model in Lyra geometry.
\section{Expressions For Hubble's Constant, Luminosity Distance, Apparent Magnitude etc.}
\subsection{Hubble's Constant}
The energy conservation Eq. $(36)$ is integrable for constant EoS parameter ($\gamma=$constant), giving rise to following expression amongst matter density $\rho$, average scale factor $a(t)=A(t)$ and the redshift $z$ of the universe
\begin{equation}\label{53}
\rho=\rho_{0}\left(\frac{A_{0}}{A}\right)^{3(1+\gamma)}=\rho_{0}(1+z)^{3(1+\gamma)} 
\end{equation}
where we have used the relation given by the Eq.$(53)$.\\
Now, using Eqs. $(38), (47), (50)$, and $(53)$, we get following expressions for Hubble's constant in terms of scale factor and redshift 
\begin{equation}\label{54}
H=\frac{H_{0}}{\sqrt{1-\frac{(1-3\gamma)(5\omega-3\omega\gamma-18\gamma+12)}{6(\omega-\omega\gamma-3\gamma+2)^{2}}}}\sqrt{\Omega_{m0}
\left(\frac{A_{0}}{A}\right)^{\frac{3\omega-3\omega\gamma^{2}+6\gamma+7}{\omega-\omega\gamma-3\gamma+2}}+\Omega_{\sigma 0}\left(\frac{A_{0}}{A}\right)^{\frac{2(3\omega-3\omega\gamma-6\gamma+5}{\omega-\omega\gamma-3\gamma+2}}+\Omega_{\beta 0}}
\end{equation}
and
\begin{equation}\label{55}
H=\frac{H_{0}}{\sqrt{1-\frac{(1-3\gamma)(5\omega-3\omega\gamma-18\gamma+12)}{6(\omega-\omega\gamma-3\gamma+2)^{2}}}}\sqrt{\Omega_{m0}
	(1+z)^{\frac{3\omega-3\omega\gamma^{2}+6\gamma+7}{\omega-\omega\gamma-3\gamma+2}}+\Omega_{\sigma 0}(1+z)^{\frac{2(3\omega-3\omega\gamma-6\gamma+5}{\omega-\omega\gamma-3\gamma+2}}+\Omega_{\beta 0}}
\end{equation}
respectively.
\subsection{Luminosity Distance}
The luminosity distance which determines flux of the source is given by
\begin{equation}\label{56}
D_{L}=A_{0}r(1+z)
\end{equation}
where $r$ is the spatial co-ordinate distance of a source. The luminosity distance for metric $(16)$ can be written as \cite{ref54}
\begin{equation}\label{57}
D_{L}=c(1+z)\int_{0}^{z}\frac{dz}{H(z)}
\end{equation}
Therefore, by using Eq.$(55)$, the luminosity distance $D_{L}$ for our model is obtained as
\begin{equation}\label{58}
D_{L}=\frac{c(1+z)\sqrt{1-\frac{(1-3\gamma)(5\omega-3\omega\gamma-18\gamma+12)}{6(\omega-\omega\gamma-3\gamma+2)^{2}}}}{H_{0}}\int_{0}^{z}\frac{dz}{\sqrt{\Omega_{m0}
	(1+z)^{\frac{3\omega-3\omega\gamma^{2}+6\gamma+7}{\omega-\omega\gamma-3\gamma+2}}+\Omega_{\sigma 0}(1+z)^{\frac{2(3\omega-3\omega\gamma-6\gamma+5}{\omega-\omega\gamma-3\gamma+2}}+\Omega_{\beta 0}}}
\end{equation}
\subsection{Apparent Magnitude}
The apparent magnitude of a source of light is related to the luminosity distance via following expression
\begin{equation}\label{59}
m=16.08+5\log_{10}\frac{H_{0}D_{L}}{0.026cMpc}
\end{equation}
Using Eq.$(58)$, we get following expression for apparent magnitude in our model
\[
m=16.08+
\]
\begin{equation}\label{60}
5\log_{10}\left((1+z)\sqrt{1-\frac{(1-3\gamma)(5\omega-3\omega\gamma-18\gamma+12)}{6(\omega-\omega\gamma-3\gamma+2)^{2}}}\int_{0}^{z}\frac{dz}{\sqrt{\Omega_{m0}
		(1+z)^{\frac{3\omega-3\omega\gamma^{2}+6\gamma+7}{\omega-\omega\gamma-3\gamma+2}}+\Omega_{\sigma 0}(1+z)^{\frac{2(3\omega-3\omega\gamma-6\gamma+5}{\omega-\omega\gamma-3\gamma+2}}+\Omega_{\beta 0}}} \right) 
\end{equation}
\subsection{Energy parameters at Present}
We consider $580$ high red shift $(0.015 \leq z \leq 1.414)$ SN Ia supernova data of observed apparent magnitudes along with their possible error from union 2.1 compilation \cite{ref55}. In our present study, we have used a technique to estimate the present values of energy parameters $\Omega_{m0}$, $\Omega_{\sigma 0}$ and $\Omega_{\beta 0}$ by comparing the theoretical and observed results with the help of $R^{2}$ formula.
\begin{equation}\label{61}
R^{2}_{SN}=1-\frac{\sum_{i=1}^{580}[(m_{i})_{ob}-(m_{i})_{th}]^{2}}{\sum_{i=1}^{580}[(m_{i})_{ob}-(m_{i})_{mean}]^{2}}
\end{equation}
Here the sums are taken over data sets of observed and theoretical values of apparent magnitude of 580 supernovae.\\

The ideal case $R^{2}=1$ occurs when the observed data and theoretical function $m(z)$ agree exactly. On the basis of maximum value of $R^{2}$, we get the best fit present values of $\Omega_{m}$, $\Omega_{\sigma}$ and $\Omega_{\beta}$ for the apparent magnitude $m(z)$ function as shown in Eq.$(60)$ which is given in Table $1$. For this, coupling constant $\omega$ is taken as $ > 40000$ and the theoretical values are calculated from Eq. $(60)$. We have found the best fit present values of $\Omega_{m}$, $\Omega_{\sigma}$ and $\Omega_{\beta}$ are $(\Omega_{m})_{0}=0.2940$, $(\Omega_{\sigma})_{0}=1.701\times 10^{-14}$ and $(\Omega_{\beta})_{0}=0.7452$ for maximum $R^{2}=0.9931$ with root mean square error (RMSE) $0.2664$ i.e. $m(z)\pm 0.2664$ and their $R^{2}$ values only $0.69\%$ far from the best one.
The Figure $3$ and Figure $4$ indicate how the observed values of luminosity distances and apparent magnitudes respectively, reach close to the theoretical graphs for $(\Omega_{m})_{0}=0.2940$, $(\Omega_{\sigma})_{0}=1.701\times 10^{-14}$, $(\Omega_{\beta})_{0}=0.7452$.
\begin{table}[H]
	\centering
	{\begin{tabular}{rrrrrrrrrrrrrrrrrrrrrrrrrrrrrrrrrrrrrrrrrrrrrrrrr@{}ccccccccccccccccccccccccccccccccccc@{}}
			\hline\hline 
			Function \vline & $\gamma$ \vline & $\omega$ \vline & $\Omega_{m0}$ \vline & $\Omega_{\sigma 0}$ \vline & $\Omega_{\beta 0}$ \vline & $H_{0}$ \vline & $R^{2}$ \vline & $RMSE$\\
            \hline\hline
			$H(z)$ \vline & $0\leq\gamma\leq\frac{1}{3}$ \vline & $49590$ \vline & $0.2991$ \vline & $2.341\times 10^{-14}$ \vline & $0.7443$ \vline & $71.27$ \vline & $0.8798$ \vline & $16.75$\\ 
			\hline
			$m(z)$ \vline & $0\leq\gamma\leq\frac{1}{3}$ \vline & $49413$ \vline & $0.2940$ \vline & $1.701\times 10^{-14}$ \vline & $0.7452$ \vline & - \vline & $0.9931$ \vline & $0.2664$\\			           
            \hline
	\end{tabular}}
	\caption{Outcomes of the $R^{2}-$test for the best fit curve of apparent magnitude $m(z)$ in Eq.$(60)$ and Hubble constant $H(z)$ in Eq. $(55)$ and Figure $4$ \& $2$. The values of coefficients $\Omega_{m}$, $\Omega_{\sigma}$, $\Omega_{\beta}$ and $\omega$ are at $95\%$ confidence of bounds.}
\end{table}
\begin{table}[H]
	\centering
	{\begin{tabular}{rrrrrrrrrrrrrrrrrrrrrrrrrrrrrrrrrrrrrrrrrrrrrrrrr@{}ccccccccccccccccccccccccccccccccccc@{}}
			\hline\hline 
			$z$  \vline & $H(z)$ \vline & $\sigma_{H}$ \vline & Reference \vline & Method\\
			\hline\hline
			$0.07$ \vline & $69$ \vline & $19.6$ \vline & Moresco M. et al. \cite{ref57} \vline & DA\\
			$0.1$ \vline & $69$ \vline & $12$ \vline & Zhang C. et al. \cite{ref58} \vline & DA\\
			$0.12$ \vline & $68.6$ \vline & $26.2$ \vline & Moresco M. et al. \cite{ref57} \vline & DA\\
			$0.17$ \vline & $83$ \vline & $8$ \vline & Zhang C. et al. \cite{ref58} \vline & DA\\
			$0.28$ \vline & $88.8$ \vline & $36.6$ \vline & Moresco M. et al. \cite{ref57} \vline & DA\\
			$0.4$ \vline & $95$ \vline & $17$ \vline & Zhang C. et al. \cite{ref58} \vline & DA\\
			$0.48$ \vline & $97$ \vline & $62$ \vline & Zhang C. et al. \cite{ref58} \vline & DA\\
			$0.593$ \vline & $104$ \vline & $13$ \vline & Moresco M. \cite{ref59} \vline & DA\\
			$0.781$ \vline & $105$ \vline & $12$ \vline & Moresco M. \cite{ref59} \vline & DA\\
			$0.875$ \vline & $125$ \vline & $17$ \vline & Moresco M. \cite{ref59} \vline & DA\\
			$0.88$ \vline & $90$ \vline & $40$ \vline & Zhang C. et al. \cite{ref58} \vline & DA\\
			$0.9$ \vline & $117$ \vline & $23$ \vline & Zhang C. et al. \cite{ref58} \vline & DA\\
			$1.037$ \vline & $154$ \vline & $20$ \vline & Moresco M. \cite{ref59} \vline & DA\\
			$1.3$ \vline & $168$ \vline & $17$ \vline & Zhang C. et al. \cite{ref58} \vline & DA\\
			$1.363$ \vline & $160$ \vline & $33.6$ \vline & Moresco M. \cite{ref59} \vline & DA\\
			$1.43$ \vline & $177$ \vline & $18$ \vline & Zhang C. et al. \cite{ref58} \vline & DA\\
			$1.53$ \vline & $140$ \vline & $14$ \vline & Zhang C. et al. \cite{ref58} \vline & DA\\
			$1.75$ \vline & $202$ \vline & $40$ \vline & Zhang C. et al. \cite{ref58} \vline & DA\\
			$1.965$ \vline & $186.5$ \vline & $50.4$ \vline & Stern D. et al. \cite{ref60} \vline & DA\\
			\hline
	\end{tabular}}
	\caption{Hubble's constant Table.}
\end{table}
\begin{figure}
	\centering
	\includegraphics[width=10cm,height=8cm,angle=0]{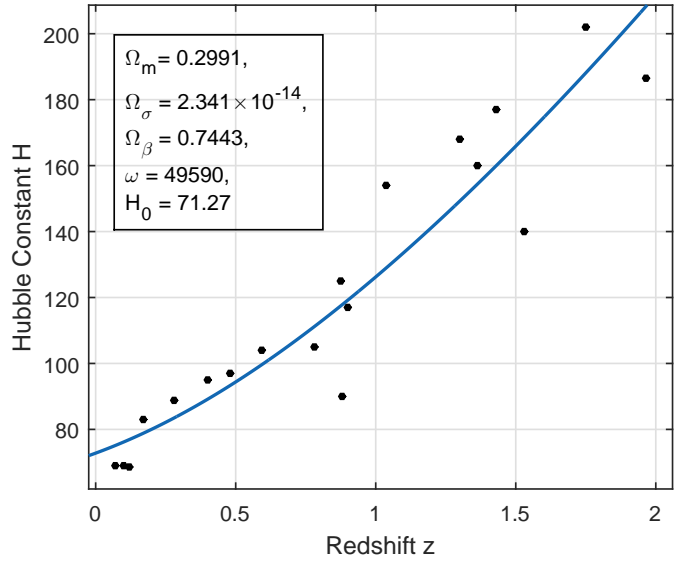}
	\caption{Hubble constant versus red shift best fit curve.}.
\end{figure}
\begin{figure}
	\centering
	\includegraphics[width=10cm,height=8cm,angle=0]{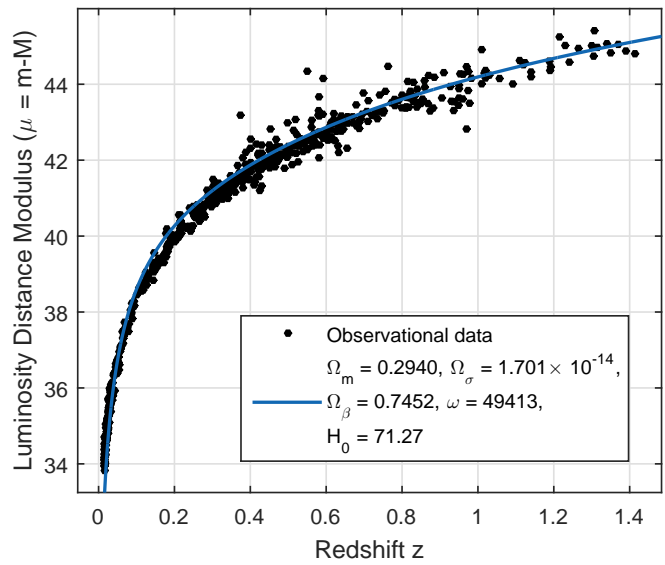}
	\caption{Luminosity distance modulus versus red shift best fit curve.}.
\end{figure}
\begin{figure}
	\centering
	\includegraphics[width=10cm,height=8cm,angle=0]{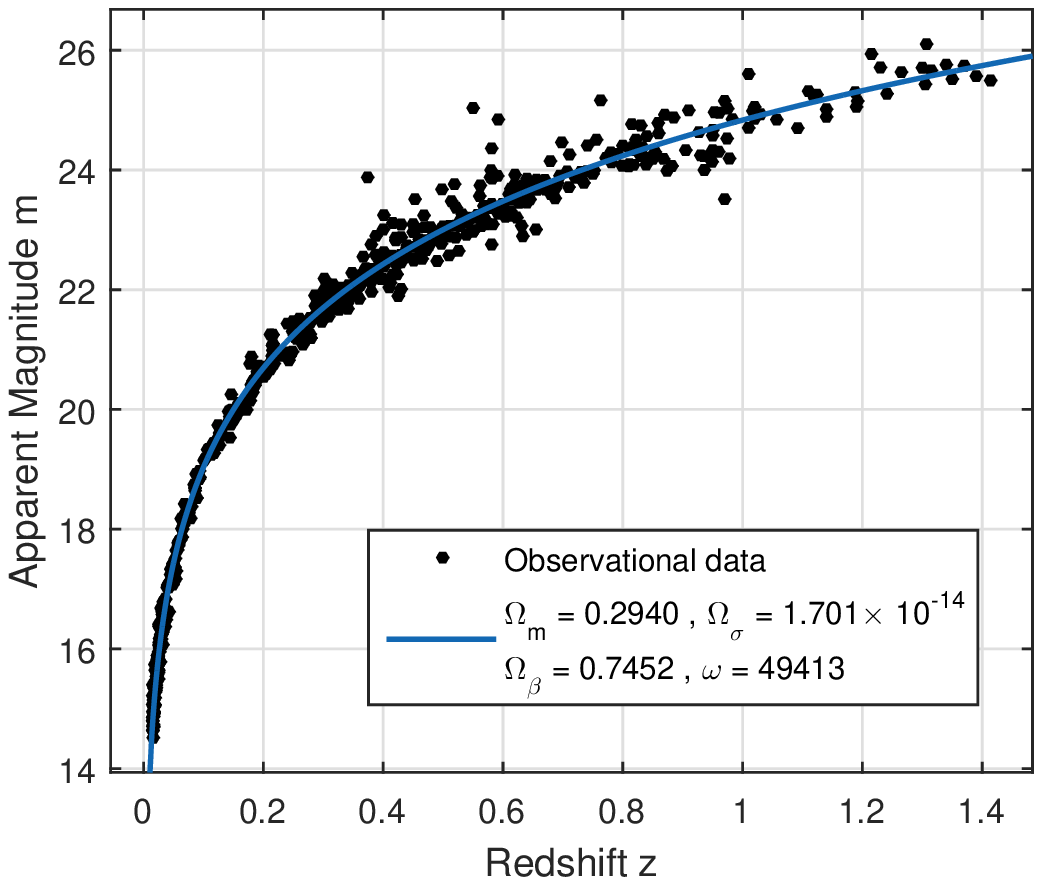}
	\caption{Apparent magnitude versus red shift best fit curve.}.
\end{figure}
\subsection{Estimation of Present Values of Hubble's Constant $H_{0}$}
We present a data set of the observed values of the Hubble parameter $H(z)$ versus the red shift $z$ with possible error in the form of Table-2. These data points were obtained by various researchers from time to time, by using different age approach.\\

In our model, Hubble's constant $H(z)$ versus red shift 'z' relation Eq. $(55)$ is reduced to
\begin{equation}\label{62}
H^{2}=(1.0003)H_{0}^{2}[0.2991(1+z)^{3}+2.341\times 10^{-14}(1+z)^{6}+0.7443]
\end{equation}
Where we have taken $(\Omega_{m})_{0}=0.2991$, $(\Omega_{\sigma})_{0}=2.341\times 10^{-14}$, $(\Omega_{\beta})_{0}=0.7443$ and the coupling constant $\omega=49590$. The Hubble Space Telescope (HST) observations of Cepheid variables \cite{ref56} provides present value of Hubble constant $H_{0}$ in the range $H_{0}=73.8\pm2.4 km/s/Mpc$. A large number of data sets of theoretical values of Hubble constant $H(z)$ versus z, corresponding to $H_{0}$ in the range $(60.45\leq H_{0} \leq 74.21)$ are obtained by using Eq. $(62)$. It should be noted that the red shift z are taken from Table-2 and each data set will consist of 19 data points.\\
In order to get the best fit theoretical data set of Hubble's constant $H(z)$ versus z, we calculate $R^{2}-$test by using following statistical formula:
\begin{equation}\label{63}
R^{2}_{SN}=1-\frac{\sum_{i=1}^{19}[(H_{i})_{ob}-(H_{i})_{th}]^{2}}{\sum_{i=1}^{19}[(H_{i})_{ob}-(H_{i})_{mean}]^{2}}
\end{equation}
Here the sums are taken over data sets of observed and theoretical values of Hubble's constants. The observed values are taken from Table-2 and theoretical values are calculated from Eq. $(55)$. Using the above $R^{2}$-test, we have found the best fit function of $H(z)$ for the Eq.$(55)$ which is mentioned in Table-1.\\

From the Table-1, one can see that the best fit value of Hubble constant $H_{0}$ is $71.27$ for maximum $R^{2}=0.8798$ with root mean square error $RMSE=16.75$ i.e. $H_{0}=71.27\pm 16.75$ and their $R^{2}$ values only $12.02\%$ far from the best one. Figure $2$ shows the dependence of Hubble's constant with red shift. Hubble's observed data points are closed to the graph corresponding to $(\Omega_{m})_{0}=0.2991$, $(\Omega_{\sigma})_{0}=2.341\times 10^{-14}$, $(\Omega_{\beta})_{0}=0.7443$. This validates the proximity of observed and theoretical values.
\begin{figure}
	\centering
	\includegraphics[width=10cm,height=8cm,angle=0]{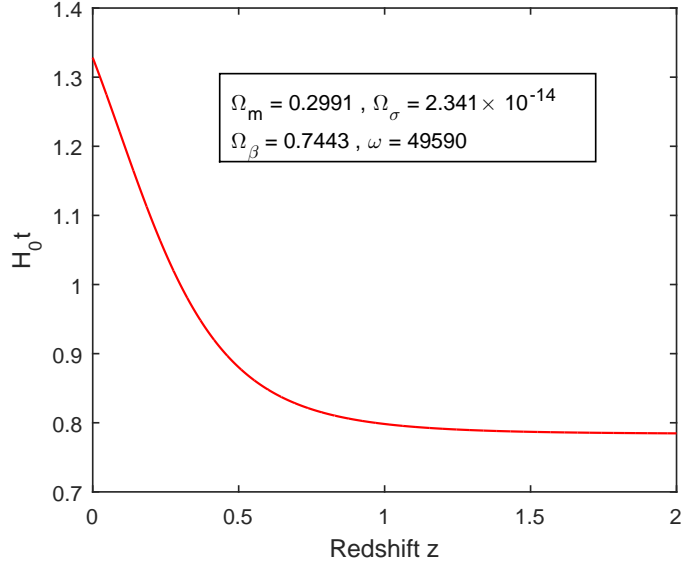}
	\caption{Plot of $H_{0}t$ versus red shift $z$.}
\end{figure}
\section{Estimation of Certain Other Physical Parameters of the Universe}
\subsection{Matter, Dark Energy And Anisotropic Energy Densities}
The matter, anisotropic energy and dark energy densities of the universe are related to the energy parameters through following equation
\begin{equation}\label{64}
\Omega_{m}=\frac{\rho_{m}}{\rho_{c}},~~~~ \Omega_{\sigma}=\frac{\rho_{\sigma}}{\rho_{c}},~~~~~\Omega_{\beta}=\frac{\rho_{\beta}}{\rho_{c}},
\end{equation}
where
\begin{equation}\label{65}
\rho_{c}=\frac{3c^{2}H^{2}}{8\pi G}=\frac{3c^{2}\phi H^{2}}{8\pi}.
\end{equation}
So,
\begin{equation}\label{66}
(\rho_{m})_{0}=(\rho_{c})_{0}(\Omega_{m})_{0}, ~~~~~(\rho_{\sigma})_{0}=(\rho_{c})_{0}(\Omega_{\sigma})_{0},~~~~(\rho_{\beta})_{0}=(\rho_{c})_{0}(\Omega_{\beta})_{0}
\end{equation}
Now the present value of $\rho_{c}$ is obtained as
\begin{equation}\label{67}
(\rho_{c})_{0}=\frac{3c^{2}H^{2}_{0}}{8\pi G}=1.88 h_{0}^{2}\times 10^{-29} gm/cm^{3}
\end{equation}
The estimated value of $h_{0}=0.7127$. Therefore, the present value of matter and dark energy densities are given by
\begin{equation}\label{68}
(\rho_{m})_{0}=0.562308 h_{0}^{2}\times 10^{-29} gm/cm^{3}
\end{equation}
\begin{equation}\label{69}
(\rho_{\sigma})_{0}=4.40108 h_{0}^{2}\times 10^{-43} gm/cm^{3}
\end{equation}
\begin{equation}\label{70}
(\rho_{\beta})_{0}=1.399284 h_{0}^{2}\times 10^{-29} gm/cm^{3}
\end{equation}
Here, we have taken  $(\Omega_{m})_{0}=0.2991$, $(\Omega_{\sigma})_{0}=2.341\times 10^{-14}$, \& $(\Omega_{\beta})_{0}=0.7443$.
General expressions for energy densities are given by
\begin{equation}\label{71}
\rho=\rho_{0}\left(\frac{a_{0}}{a}\right)^{3(1+\gamma)}=\rho_{0}(1+z)^{3(1+\gamma)} 
\end{equation}
\begin{equation}\label{72}
\rho_{\sigma}=\frac{k^{2}c^{2}\phi^{3}}{8\pi}\left(\frac{a_{0}}{a}\right)^{6}=\frac{k^{2}c^{2}\phi^{3}}{8\pi}(1+z)^{6}
\end{equation}
and
\begin{equation}\label{73}
\rho_{\beta}=(\rho_{c}) \Omega_{\beta}=\frac{3c^{2}\beta^{2}}{32\pi}\phi
\end{equation}
From above, we observe that the current matter and dark energy densities are very close to the values predicted by the various surveys described in the introduction.
\subsection{Age of The Universe}
By using the standard formula 
\begin{equation}\nonumber
t=\int_{0}^{t}dt=\int_{0}^{A}\frac{dA}{AH}
\end{equation}
we obtain the values of $t$ in terms of scale factor and redshift respectively:
\begin{equation}\label{74}
t=\int_{0}^{A}\frac{\sqrt{1-\frac{(1-3\gamma)(5\omega-3\omega\gamma-18\gamma+12)}{6(\omega-\omega\gamma-3\gamma
	+2)^{2}}}dA}{AH_{0}\sqrt{\Omega_{m0}
\left(\frac{A_{0}}{A}\right)^{\frac{3\omega-3\omega\gamma^{2}+6\gamma+7}{\omega-\omega\gamma-3\gamma+2}}+\Omega_{\sigma 0}\left(\frac{A_{0}}{A}\right)^{\frac{2(3\omega-3\omega\gamma-6\gamma+5}{\omega-\omega\gamma-3\gamma+2}}+\Omega_{\beta 0}}}
\end{equation}
\begin{equation}\label{75}
t=\int_{0}^{z}\frac{\sqrt{1-\frac{(1-3\gamma)(5\omega-3\omega\gamma-18\gamma+12)}{6(\omega-\omega\gamma-3\gamma
			+2)^{2}}}dz}{(1+z)H_{0}\sqrt{\Omega_{m0}
		(1+z)^{\frac{3\omega-3\omega\gamma^{2}+6\gamma+7}{\omega-\omega\gamma-3\gamma+2}}+\Omega_{\sigma 0}(1+z)^{\frac{2(3\omega-3\omega\gamma-6\gamma+5}{\omega-\omega\gamma-3\gamma+2}}+\Omega_{\beta 0}}}
\end{equation}
For $\omega=49590$, $(\Omega_{m})_{0}=0.2991$, $(\Omega_{\sigma})_{0}=2.341\times 10^{-14}$, \& $(\Omega_{\beta})_{0}=0.7443$, Eq.$(75)$ gives $t_{0}\to 1.3282H_{0}^{-1}$ for high redshift. This means that the present age of the universe is $t_{0}=18.23^{+5.60}_{-3.47}$ Gyrs as per our model. From WMAP data, the empirical value of present age of universe is $13.73\pm 0.13$Gyrs which is closed to present age of universe, estimated by us in this paper.\\
\begin{figure}
	\centering
	\includegraphics[width=10cm,height=8cm,angle=0]{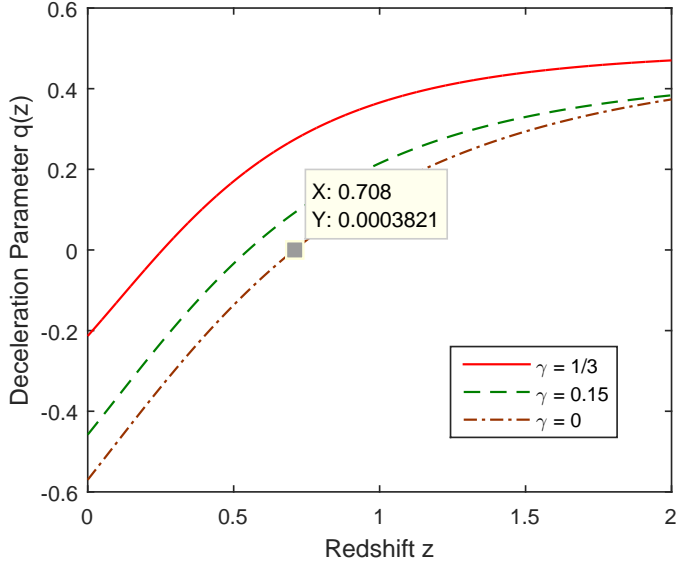}
	\caption{Variation of deceleration parameter versus red shift.}
\end{figure}
Figure $5$ shows the variation of time over redshift. At $z=0$ the value of $H_{0}t_{0}=1.3282$. This provides present age of the universe. This also indicated the consistency with recent observations.

\subsection{Deceleration Parameter}
From Eqs. $(41),~(44)$ and $(46)$, we obtain the expressions for DP as
\begin{equation}\label{76}
q=-\frac{\omega+\omega\gamma+1}{2\omega\gamma+3\gamma-2\omega-3}+\frac{(\omega-3\omega\gamma+2)(1-3\gamma)}{2(\omega-\omega\gamma-3\gamma+2)(2\omega\gamma+3\gamma-2\omega-3)}+\frac{3(\omega-\omega\gamma-3\gamma+2)}{2\omega\gamma+3\gamma-2\omega-3}(\gamma~\Omega_{m}+\Omega_{\sigma}+\Omega_{\beta})
\end{equation}
Using Eqs. $(38),~(46),~(54)$ and $(55)$ in Eq.$(76)$, we get following expression for deceleration parameter
\[
q=-\frac{\omega+\omega\gamma+1}{2\omega\gamma+3\gamma-2\omega-3}+\frac{(\omega-3\omega\gamma+2)(1-3\gamma)}{2(\omega-\omega\gamma-3\gamma+2)(2\omega\gamma+3\gamma-2\omega-3)}
\]
\begin{equation}\label{77}
+\frac{3(\omega-\omega\gamma-3\gamma+2)}{2\omega\gamma+3\gamma-2\omega-3}\frac{\left( {1-\frac{(1-3\gamma)(5\omega-3\omega\gamma-18\gamma+12)}{6(\omega-\omega\gamma-3\gamma+2)^{2}}}\right) \left[\gamma\Omega_{m0}
	\left(\frac{A_{0}}{A}\right)^{\frac{3\omega-3\omega\gamma^{2}+6\gamma+7}{\omega-\omega\gamma-3\gamma+2}}+\Omega_{\sigma 0}\left(\frac{A_{0}}{A}\right)^{\frac{2(3\omega-3\omega\gamma-6\gamma+5)}{\omega-\omega\gamma-3\gamma+2}}+\Omega_{\beta 0}\right]}{\left[ \Omega_{m0}
	\left(\frac{A_{0}}{A}\right)^{\frac{3\omega-3\omega\gamma^{2}+6\gamma+7}{\omega-\omega\gamma-3\gamma+2}}+\Omega_{\sigma 0}\left(\frac{A_{0}}{A}\right)^{\frac{2(3\omega-3\omega\gamma-6\gamma+5)}{\omega-\omega\gamma-3\gamma+2}}+\Omega_{\beta 0}\right]}
\end{equation}
In terms of red shift, $q$ is given by
\[
q=-\frac{\omega+\omega\gamma+1}{2\omega\gamma+3\gamma-2\omega-3}+\frac{(\omega-3\omega\gamma+2)(1-3\gamma)}{2(\omega-\omega\gamma-3\gamma+2)(2\omega\gamma+3\gamma-2\omega-3)}
\]
\begin{equation}\label{78}
+\frac{3(\omega-\omega\gamma-3\gamma+2)}{2\omega\gamma+3\gamma-2\omega-3}\frac{\left( {1-\frac{(1-3\gamma)(5\omega-3\omega\gamma-18\gamma+12)}{6(\omega-\omega\gamma-3\gamma+2)^{2}}}\right) \left[\gamma\Omega_{m0}
	(1+z)^{\frac{3\omega-3\omega\gamma^{2}+6\gamma+7}{\omega-\omega\gamma-3\gamma+2}}+\Omega_{\sigma 0}(1+z)^{\frac{2(3\omega-3\omega\gamma-6\gamma+5)}{\omega-\omega\gamma-3\gamma+2}}+\Omega_{\beta 0}\right]}{\left[ \Omega_{m0}
	(1+z)^{\frac{3\omega-3\omega\gamma^{2}+6\gamma+7}{\omega-\omega\gamma-3\gamma+2}}+\Omega_{\sigma 0}(1+z)^{\frac{2(3\omega-3\omega\gamma-6\gamma+5)}{\omega-\omega\gamma-3\gamma+2}}+\Omega_{\beta 0}\right]}
\end{equation}
At the present phase $(z=0)$ of the universe is accelerating $q\leq 0$ i.e. $\frac{\ddot{a}}{a}\geq 0$, so we must have\\
\begin{equation}\label{79}
\Omega_{\beta 0}\geq \left[\frac{-3\gamma^{2}+4\gamma+1}{2(\gamma+2)}+\frac{3(\gamma-1)[-24\gamma^{2}+8\gamma+4-\omega(9\gamma^{3}-21\gamma^{2}+7\gamma-3)]}{2(\gamma+2)[(12\gamma^{2}-36\gamma+24)\omega^{2}+(45\gamma^{2}-126\gamma+73)\omega-60\gamma+44]}\right]\Omega_{m0}-\Omega_{\sigma 0} 
\end{equation}
For $\omega=49590$ and $\Omega_{m}=0.2991$, $\Omega_{\sigma}=2.341\times 10^{-14}$, $0\leq \gamma < \frac{1}{3}$ the minimum value of 
$\Omega_{\beta 0}$ is given by $\Omega_{\beta 0}\geq 0.3152$ which is consistent with the present observed value of $\Omega_{\beta 0}=0.7443$.\\
Putting $z=0$ in Eq. $(79)$, the present value of deceleration parameter is obtained as
\begin{equation}\label{80}
q_{0}=-0.57.
\end{equation}
The Eq. $(79)$ also provides
\begin{equation}\label{81}
z_{c}\approx 0.708 ~~~~at ~~~~q=0
\end{equation}
Therefore, the universe attains to the accelerating phase when $z < z_{c}$.\\

Converting redshift into time from Eq. $(81)$, the value of $z_{c}$ is reduced to
\begin{equation}\label{82}
z_{c}=0.708\sim 0.8258 H_{0}^{-1} Gyrs\sim 11.33 Gyrs
\end{equation}
So, the acceleration must have begun in the past at $11.33 Gyrs$. The Figure $6$ shows how deceleration parameter increases from negative to positive over red shift which means that in the past universe was decelerating and at a instant $z_{c}\cong708$, it became stationary there after it goes on accelerating.
\subsection{Shear Scalar}
The shear scalar is given by 
\begin{equation}\label{83}
\sigma^{2}=\frac{1}{2}\sigma_{ij}\sigma^{ij}
\end{equation}
where
\begin{equation}\label{84}
\sigma_{ij}=u_{i;j}-\theta(g_{ij}-u_{i}u_{j})
\end{equation}
In our model
\begin{equation}\label{85}
\sigma^{2}=\frac{\dot{d}^{2}}{d^{2}}=\frac{k^{2}\phi^{2}}{A^{6}}=(\Omega_{\sigma})_{0}H_{0}^{2}(1+z)^{\frac{2(3\omega-3\omega\gamma-6\gamma+5)}{\omega-\omega\gamma-3\gamma+2}}
\end{equation}
From Eq. $(85)$, it is clear that shear scalar vanishes as $A\to\infty$.
\subsection{Relative Anisotropy}
The relative anisotropy is given by
\begin{equation}\label{86}
\frac{\sigma^{2}}{\rho_{m}}=\frac{3\Omega_{\sigma 0}H_{0}^{2}(1+z)^{\frac{2(3\omega-3\omega\gamma-6\gamma+5)}{\omega-\omega\gamma-3\gamma+2}}}{(\rho_{c})_{0}\Omega_{m0}}
\end{equation}
This follows the same pattern as shear scalar. This means that relative anisotropy decreases over scale factor i.e. time.
\section{Discussion of Results}
For the viability of a cosmological model, it is necessary that it should be consistent with recent observational data. From the analysis of various observational data, some of the basic observational parameters $\Omega_{0}$, $q_{0}$, and $H_{0}$ read as total density parameter, deceleration parameter, and Hubble constant respectively are evaluated. On the basis of these, it has been established now that the expansion of the Universe is in accelerating phase at present. Since there are not enough matter or radiation to lead this accelerated expansion, theoreticians attributed this to a mysterious form of energy present in the universe and termed it as dark energy (DE). The earlier discarded cosmological constant $\Lambda$ is reintroduced to incorporate DE. But the question is, what is the actual nature of this dark energy and from where the $\Lambda$-term comes$?$ To answer this, various researchers using different approaches, proposed different models, but despite of these, it is still a mystery for the researchers.\\
\par In this order, in the present paper, we have proposed a cosmological model of the universe in Bianchi type-I spacetime in the context of an amalgamation of BD theory and Lyra geometry. The Lyra geometry establishes a term like cosmological constant $\Lambda$, which, otherwise is simply added in Einstein field equation by Einstein and in some other theories. The BD theory is involved due to make the model consistent with Mach's principle, since the Einstein's GR and also its Lyra's modification are inconsistent with it. Here, we have solved the field equations $(17)$ to $(21)$ taking EoS parameter $\gamma=$ constant. We have defined the energy density parameters $\Omega_{m}$ as matter energy density parameter, $\Omega_{\sigma}$ as curvature energy density parameter and $\Omega_{\beta}$ as dark energy density parameter. Since the observational data are available in the form of apparent magnitude and Hubble constant $H$ with redshift $z$, therefore, we have derived the expressions for the apparent magnitude $m(z)$ and Hubble constant $H(z)$ in terms of energy parameters $\Omega_{m}$, $\Omega_{\sigma}$, $\Omega_{\beta}$ and redshift $z$ as in Eqs.$(60)$ and $(55)$ respectively.\\
\par For mapping the theoretical model to observed universe model we have fitted the curve of Hubble constant $H(z)$ and apparent magnitude $m(z)$ using $R^{2}$-test formula. On the basis of maximum $R^{2}$ value, we obtained best fit curve for $m(z)$ to $580$ data of union $2.1$ compilation data of SNe Ia observations \cite{ref55}, with coefficients $\Omega_{m}=0.2940$, $\Omega_{\sigma}=1.701\times 10^{-14}$, $\Omega_{\beta}=0.7452$ and $\omega=49413$ with $95\%$ confidence of bounds and maximum $R^{2}=0.9931$. On the other hand, we have found the best fit curve for $H(z)$ with coefficients $\Omega_{m}=0.2991$, $\Omega_{\sigma}=2.341 \times 10^{-14}$, $\Omega_{\beta}=0.7443$, $H_{0}=71.27$ and $\omega=49590$ with maximum $R^{2}=0.8798$. From these two fittings (mentioned in Table-1) with different source of data sets, one can see that the values of density parameters $\Omega_{m}$, $\Omega_{\sigma}$, $\Omega_{\beta}$ are approximately compatible to each other. If we compare these values with the values obtained from analysis of observational data sets, we find that these are very close to them. From Eq.$(47)$, one can see that total energy density parameter is greater than unity, which means our model supports an open universe.\\
\par In section-$5.3$, we have obtained the expression for the deceleration parameter $q(z)$ in Eq.$(78)$ in terms of density parameters and redshift $z$. Figure $6$ represents the plot of DP versus redshift $z$ with coefficients $\Omega_{m}=0.2991$, $\Omega_{\sigma}=2.341\time 10^{-14}$, $\Omega_{\beta}=0.7443$ and $\omega=49590$ for $\gamma=0.3,~0.15,~0$. From the Figure $6$, one can see that $q(z)$ is an increasing function of redshift $z$. A non-zero deceleration parameter $q(z)$ depicts the expansion phase of the Universe (decelerating or accelerating, as its value is positive or negative respectively). Here, from the figure $6$, one can see that signature of $q$ changes at $z_{c}\approx 0.7080$ and this point is called transition point of the model. For $z > z_{c}=0.708$, $q > 0$ indicating the universe in decelerating phase and for $z < z_{c}=0.708$, $q < 0$ indicating the expansion is in accelerating phase. It means the expansion of the universe is entered into the accelerating phase at $z\approx 0.708$ (see Eq.(81)) which is equivalent to the age of $11.33 ~ Gyrs$ (see Eq.(82)). The direct empirical evidence for the transition from past deceleration to present acceleration is provided by SNe type Ia measurements. In their preliminary analysis it was found that the SNe data favor recent acceleration ($z < 0.5$) and past deceleration ($z > 0.5$). More recently, the High-z Supernova Search (HZSNS) team have obtained $z_{t} = 0.46 \pm 0.13$ at
$(1\;\sigma)$ c.l. \cite{ref62} in 2004 which has been further improved to  $z_{t} = 0.43 \pm 0.07$ at $(1\;\sigma)$ c.l. \cite{ref62} in 2007. The Supernova Legacy Survey (SNLS) (Astier et al. \cite{ref63}), as well as the one recently compiled by Davis et al. 
\cite{ref64}, ( in better agreement with the flat $\Lambda$CDM model $z_{t} = (2\Omega_{\Lambda}/\Omega_{m})^{\frac{1}{3}} - 1 \sim 0.66$), yield a transition redshift $z_{t} \sim 0.6 (1\; \sigma)$.  Another limit is $0.60 \leq z_{t} \leq 1.18$ ($2\sigma$, joint analysis) \cite{ref65}. Also, in Eq.(80), we see that at $z=0$ (denoting the present stage of the universe) $q=-0.57$, which is in good agreement with recent observations.\\

\par In section-$5.2$, we have estimated the present age of the universe, which comes out to be  $t_{0}\to 1.3282H_{0}^{-1}$, for $\omega=49590$, $(\Omega_{m})_{0}=0.2991$, $(\Omega_{\sigma})_{0}=2.341\times 10^{-14}$, \& $(\Omega_{\beta})_{0}=0.7443$, which is closed to the empirical value from WMAP data.\\
	  
\par So, we can say that, these estimations establish the viability of our model.\\
\section{Conclusion}
We summarize our results by presenting Table-3 which displays the values of cosmological parameters at present obtained by us.
\begin{table}[H]
	\centering
	{\begin{tabular}{rrrrrrrrrrrrrrrrrrrr@{}cccccccccccccccccccccccccccc@{}}
			\hline\hline 
			Cosmological Parameters \vline & Values at Present\\
			\hline\hline 
             BD coupling constant $\omega$ \vline & $49590$ \\
             Matter energy parameter $\Omega_{m0}$ \vline & $0.2991$\\
             Dark energy parameter $\Omega_{\beta 0}$ \vline & 0.$7443$\\
             Anisotropic energy parameter $\Omega_{\sigma 0}$ \vline & $2.341\times 10^{-14}$\\
             Hubble's constant $H_{0}$ \vline & $71.27$\\
             Deceleration parameter $q_{0}$ \vline & $-0.57$\\
             Matter energy density $\rho_{m0}$ \vline & $0.562308 h_{0}^{2}\times 10^{-29} gm/cm^{3}$ \\
             Dark energy density $\rho_{\beta0}$ \vline & $1.39928 h_{0}^{2}\times 10^{-29} gm/cm^{3}$\\
             Anisotropic energy density $\rho_{\sigma0}$ \vline & $4.40108 h_{0}^{2}\times 10^{-43} gm/cm^{3}$\\
             Age of the Universe $t_{0}$ \vline & $18.23^{+5.60}_{-3.47}$ Gyrs\\
			\hline
	\end{tabular}}
	\caption{Cosmological parameters at present for $0\leq\gamma < \frac{1}{3}$.}
\end{table}
 We have found the following main features of the model:

\begin{itemize}
	\item The derived model is an anisotropic Bianchi type-I universe which tends to isotropic flat $\Lambda$CDM model at the late time because of $\Omega_{\sigma 0}\approx 0$ and shear scalar $\sigma^{2}\to 0$ as $A\to \infty$.
	\item The values of density parameters $\Omega_{m0}$, $\Omega_{\sigma 0}$ and $\Omega_{\beta 0}$ obtained are very close to  $H(z)$ and SNe Ia data. It declares that the viability of the model.
	\item The present values of various physical parameters calculated and presented in the Table-3 are in good agreement with the recent observations.
	\item The deceleration parameter shows signature flipping (from positive to negative with decreasing redshift, see in Figure $6$) i.e. transition phase at $z_{c}\approx 0.7080$ which is in good agreement with various relativistic models and cosmological surveys.
	\item We have found that the acceleration would have begun in the past at $11.33^{+3.48}_{-3.16} Gyrs$.
	\item The Lyra geometry with constant displacement vector removes the cosmological constant $\Lambda$-term problem naturally.
	\item The present universe is dominated by scalar field $\phi$ and it is the responsible candidate for the present behaviour of the Universe.
\end{itemize}

 These results are in good agreement with the various observational results described in the introduction. In the present model, the energy parameter $\Omega_{\beta}$ behaves like dark energy parameter $\Omega_{\Lambda}$. The model creates more interest in researchers to study the behavior of gauge function $\beta$ and scalar field $\phi$ and their coupling in formulation of the universe model.
	\section*{Acknowledgment}
	The authors are thankful to IASE (Deemed to be University), Sardarshahar, Rajsthan, and GLA University, Mathura, Uttar Pradesh, India for providing facilities and support where part of this work is carried out.
		
\end{document}